\documentclass[aps,prl,superscriptaddress,floatfix,reprint]{revtex4-1}  

\usepackage{graphicx}
\usepackage{bm}
\usepackage{amssymb}
\usepackage{amsmath}
\usepackage{float}
\usepackage{pstricks} 
\usepackage{color}
\usepackage{bbold}
\usepackage{xr} 
\usepackage{tikz}
\usetikzlibrary{matrix}

\graphicspath{{figures/}}

\begin{document}

\title{An Optimal Design for Universal Multiport Interferometers}

\author{William~R.~Clements}
\email{william.clements@physics.ox.ac.uk}
\affiliation{Clarendon Laboratory, Department of Physics, University of Oxford, Oxford OX1 3PU, UK}

\author{Peter C. Humphreys}
\affiliation{Clarendon Laboratory, Department of Physics, University of Oxford, Oxford OX1 3PU, UK}

\author{Benjamin J. Metcalf}
\affiliation{Clarendon Laboratory, Department of Physics, University of Oxford, Oxford OX1 3PU, UK}

\author{W. Steven Kolthammer}
\affiliation{Clarendon Laboratory, Department of Physics, University of Oxford, Oxford OX1 3PU, UK}

\author{Ian A. Walmsley}
\affiliation{Clarendon Laboratory, Department of Physics, University of Oxford, Oxford OX1 3PU, UK}

\date{\today}

\begin{abstract}{Universal multiport interferometers, which can be programmed to implement any linear transformation between multiple channels, are emerging as a powerful tool for both classical and quantum photonics. These interferometers are typically composed of a regular mesh of beam splitters and phase shifters, allowing for straightforward fabrication using integrated photonic architectures and ready scalability. The current, standard design for universal multiport interferometers is based on work by Reck \textit{et al} (Phys. Rev. Lett. \textbf{73}, 58, 1994). We demonstrate a new design for universal multiport interferometers based on an alternative arrangement of beam splitters and phase shifters, which outperforms that by Reck \textit{et al}. Our design occupies half the physical footprint of the Reck design and is significantly more robust to optical losses.}
\end{abstract}

\maketitle

Reconfigurable universal multiport interferometers, which can implement any linear transformation between several optical channels, are emerging as a powerful tool for fields such as microwave photonics \cite{ref1,capmany2016programmable}, optical networking \cite{chen2011photonic,stabile2016integrated}, and quantum photonics \cite{reck1994experimental,carolan2015universal}. Such devices are typically built using planar meshes of beam splitters, which are easy to fabricate and to individually control, as recent demonstrations of large, yet non-universal, interferometers have shown \cite{tanizawa2015ultra, harris2015bosonic}. While it had been known for some time that useful operations could be performed by such meshes \cite{gaylor1990integrated}, the seminal work by Reck \textit{et al} \cite{reck1994experimental} demonstrated that a specific triangular mesh of $2 \times 2$ beam splitters and phase shifters could be programmed, using a simple analytical method, to implement any unitary transformation between a set of optical channels. Continued interest in universal multiport interferometers for classical and quantum applications has led to new applications and programming procedures for the same interferometer design \cite{miller2013self,miller2015perfect}. Recent demonstrations of universal multiport interferometers are based on this design, and have achieved transformations between up to six channels \cite{carolan2015universal}. 

\begin{figure*}
\center
\includegraphics[width=15cm,angle=0]{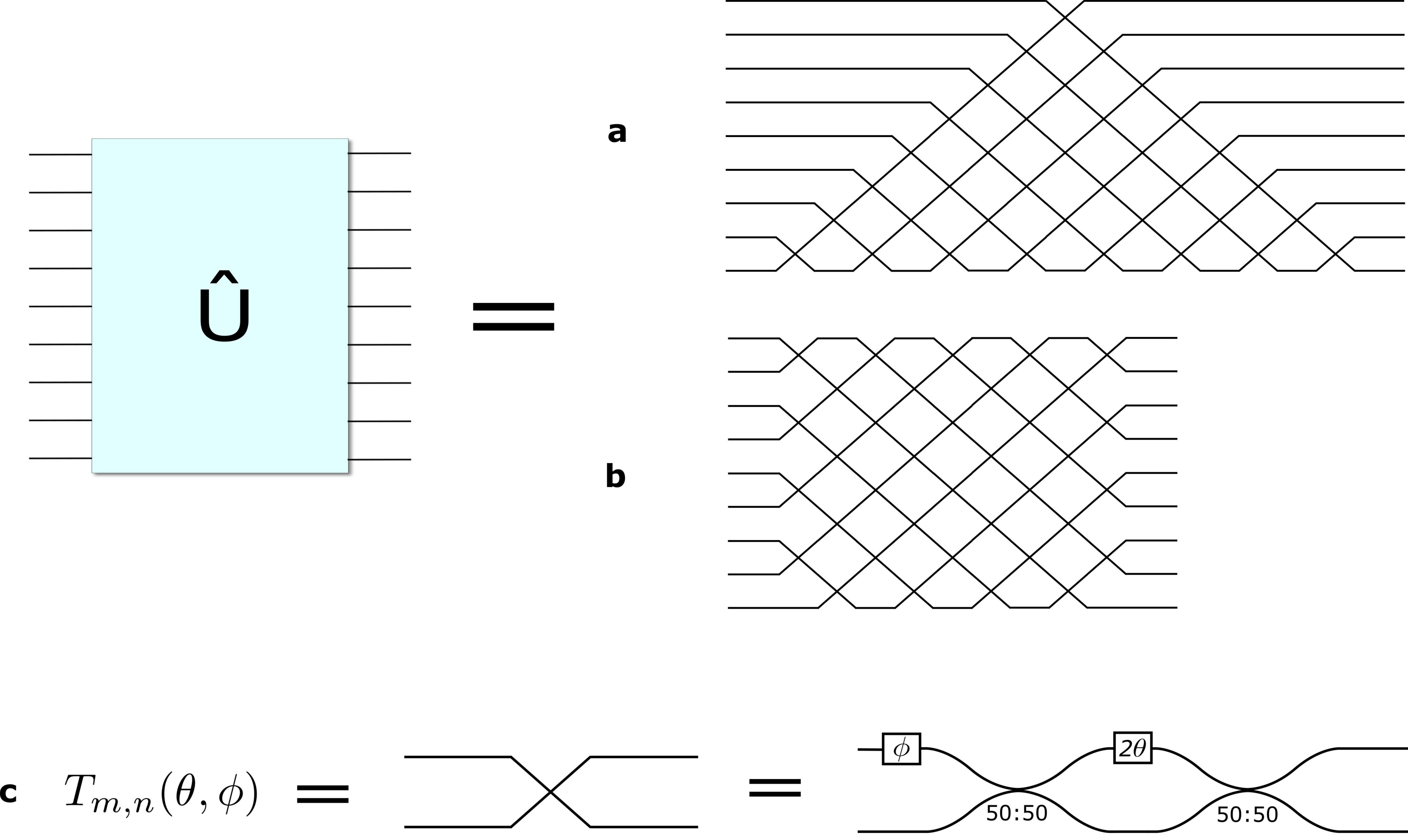}\\
\caption{A universal $N$-mode multiport interferometer (shown here for $N=9$) can be implemented using a mesh of $N(N-1)/2$ beam splitters such as a) the one proposed by Reck or b) the one that we demonstrate in this paper. As shown in c), a line corresponds to an optical mode, and crossings between two modes correspond to a variable beam splitter described by a $T_{m,n}(\theta,\phi)$ matrix, which can be implemented by a Mach-Zehnder interferometer consisting of two 50:50 directional couplers, preceded by a phase shift at one input port. Although the total number of beam splitters in both interferometers is identical, our scheme is clearly more compact, and therefore suffers less propagation loss. This compactness stems from the fact that each mode crosses its nearest neighbor at the first possible occasion, in contrast to the Reck scheme where the top modes must propagate for some distance before interacting with other modes. Furthermore, the high symmetry inherent to our design improves the loss tolerance of the interferometer, as we show in the main text.}
\label{fig1}
\end{figure*}

In this paper, we demonstrate a new design for universal multiport interferometers, based on an alternative arrangement of beam splitters and phase shifters (figure \ref{fig1}), that outperforms the design by Reck \textit{et al} (referred to as the ``Reck" design in the following). Our design occupies half the physical footprint of the Reck design and is significantly more robust to optical losses. Our finding is based on a new mathematical decomposition of a unitary matrix. We use this decomposition both to prove universality of the design and to construct an efficient algorithm to program interferometers based on it. In the following, we first provide an overview of both the Reck design and of our new design, and discuss some advantages of the latter. We then explain the general principles of our decomposition procedure using a $5 \times 5$ universal transformation as an example. Finally, we quantitatively compare the loss tolerance of our design to that of the Reck design.

\subsection{Background}

An ideal, lossless multiport interferometer between $N$ channels performs an optical transformation which can be described by an $N \times N$ unitary scattering matrix $U$ acting on electric fields as $\bm{E_{out}}=U\bm{E_{in}}$. Equivalently, in quantum optics, $U$ describes the transformation of the annihilation operators of the input modes to those of the output modes.

Within this framework, the following transformation between channels $m$ and $n$ ($m=n-1$):

\begin{equation}
T_{m,n}(\theta,\phi) =
\begin{tikzpicture}[baseline=(current bounding box.center)]
\matrix (m) [matrix of math nodes,nodes in empty cells,right delimiter={]},left delimiter={[} ]{
1  & 0 & & & & & & 0  \\
0  & 1 & & & & & &  \\
 & & & & & & &    \\
   & & & e^{i\phi} \cos \theta& - \sin \theta   & & &  \\
  & & & e^{i\phi} \sin \theta & \cos \theta& & & \\
  & & & & & & & \\
  & & & & & & 1 & 0 \\
0 & & & & & & 0 & 1\\
} ;
\draw[loosely dotted] (m-2-2)-- (m-4-4);
\draw[loosely dotted] (m-5-5)-- (m-7-7);
\draw[loosely dotted] (m-2-1)-- (m-8-1);
\draw[loosely dotted] (m-1-8)-- (m-7-8);
\draw[loosely dotted] (m-1-2)-- (m-1-8);
\draw[loosely dotted] (m-8-1)-- (m-8-7);
\end{tikzpicture}
\end{equation}

\noindent
corresponds to a lossless beam splitter between channels $m$ and $n$ with reflectivity $\cos\theta$ ($\theta \in [0,\pi/2]$), and a phase shift $\phi$ ($\phi \in [0,2\pi]$) at input $m$. In the following, we will generally omit the explicit dependence of these $T_{m,n}(\theta,\phi)$ matrices on $\theta$ and $\phi$ for notational simplicity.

Both our scheme and the scheme by Reck \textit{et al} are based on analytical methods of decomposing the $U$ matrix into a product of $T_{m,n}$ matrices. Specifically, these schemes provide an explicit algorithm for writing any unitary matrix $U$ as:

\begin{align}
U = D \left(\prod_{(m,n) \in S} T_{m,n} \right)
\label{generaldecompose}
\end{align}

\noindent
where $S$ defines a specific ordered sequence of two-mode transformations, and where $D$ is a diagonal matrix with complex elements with modulus equal to one on the diagonal. A physical interferometer composed of beam splitters and phase shifters in the configuration defined by $S$, with values defined by the $\theta$ and $\phi$ in the $T_{m,n}$ matrices, will therefore implement transformation $U$. We note that $D$ is physically irrelevant for most applications, but can be implemented in an interferometer nonetheless by phase shifts on all individual channels at the output of an interferometer.

The formalism developed here for unitary transformations describing lossless $N \times N$ interferometers can be extended to include any $M \times N$ linear (non-unitary) transformation. Indeed, it has been noted that any $M \times N$ linear transformation, with for example $M \leq N$ (resp. $M \geq N$), can be straightforwardly embedded within a $2N \times 2N$ (resp. $2M \times 2M$) unitary transformation \cite{miller2013self, alber2003quantum} to within a scaling factor. Furthermore, realistic, lossy interferometers can also be included in our formalism simply by rescaling $U$ by a loss factor, as we explain later. Therefore, our design for universal multiport interferometers, as well as that by Reck \textit{et al}, can be used to implement any linear transformation, to within a scaling factor, on any number of input and output channels.

\subsection{Overview of the two designs}

Schematic views of the Reck design and of our design are presented in figure \ref{fig1}. Figure 1a presents the Reck design, in which the matrix decomposition method determines a sequence $S$ that corresponds to to a triangular mesh of beam splitters. Figure 1b presents our design, in which every mode crosses its nearest neighbor at the first possible occasion. Our design is more compact and symmetric than the Reck design. We note that both interferometers use the same, minimal number $N(N-1)/2$ of beam splitters to implement an $N \times N$ interferometer \cite{reck1994experimental}.

We define the depth of an interferometer to be the longest path through the interferometer, enumerated by counting the number of beam splitters traversed by that path. It is important to minimize the optical depth of an interferometer because larger interferometers can then be built within a given area, which is an important constraint for fabrication of planar waveguide interferometers. Furthermore, propagation losses are reduced for an interferometer with smaller depth. It is easy to see that our design has the minimal possible optical depth, since every channel crosses its nearest neighbor at the first possible occasion. Specifically, for an $N \times N$ interferometer, the Reck design has an optical depth of $2N-3$, whereas our design has an optical depth of $N$. To illustrate this, the longest path through the interferometer shown in figure \ref{fig1}a follows the edges of the triangle and crosses $2N-3=15$ beam splitters, whereas the longest paths through the interferometer in figure \ref{fig1}b cross $N=9$ beam splitters.

The increased symmetry of our design also leads to significantly better loss tolerance, and is discussed later. 

\subsection{Decomposition method}

\begin{figure*}
\center
\includegraphics[width=17.5cm,angle=0]{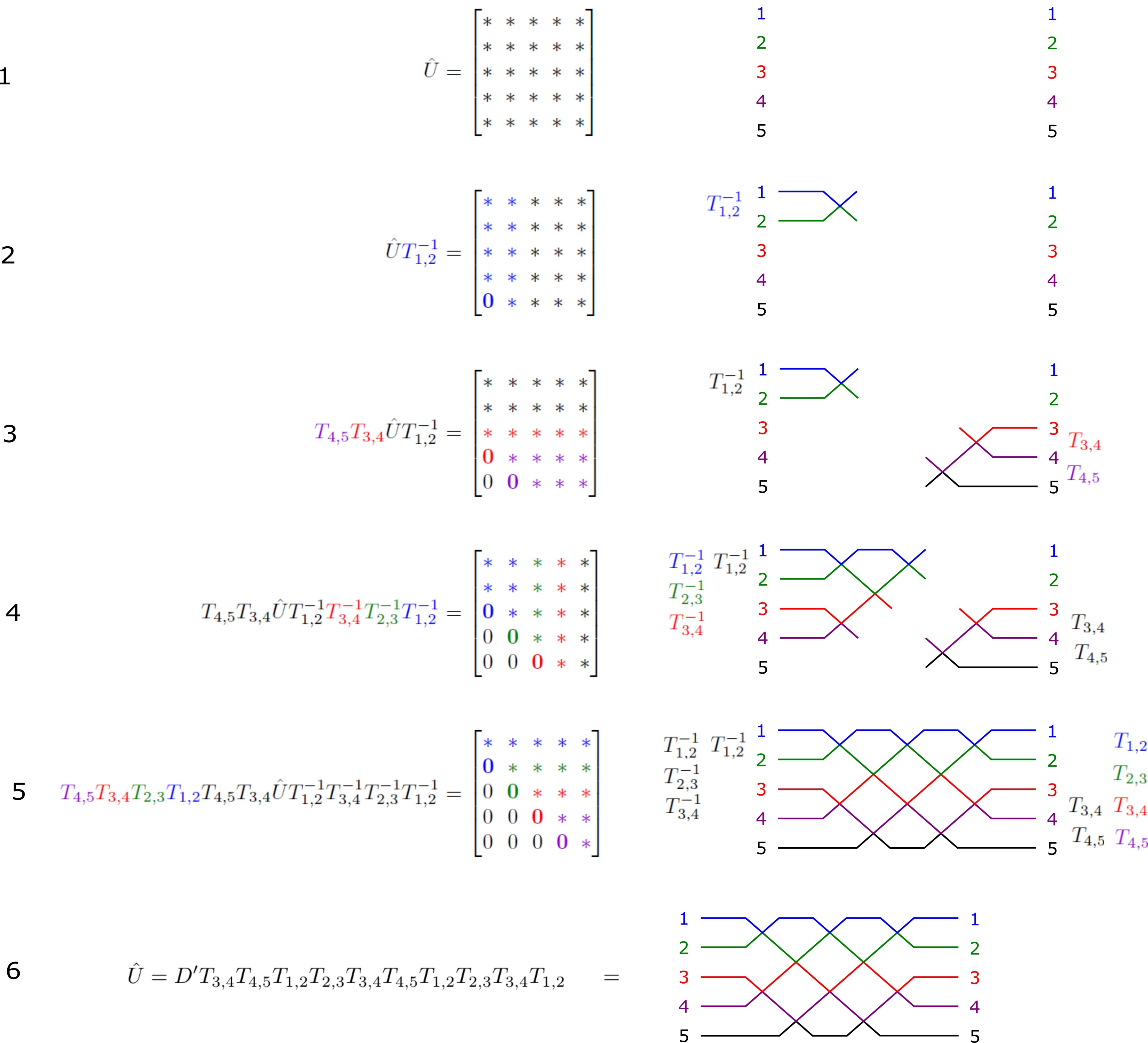}\\
\caption{Illustration of the algorithm for programming a universal multiport interferometer, for a $5 \times 5$ interferometer. The left-hand side presents our decomposition procedure, and the right-hand side shows how our decomposition corresponds to building up the corresponding interferometer. 1) We start with any random unitary matrix $U$, and a blank interferometer. 2) We first null the bottom left element of $U$ with a $T^{-1}_{1,2}$ matrix, which causes the first two columns of $U$ to mix. This corresponds to adding the top-left beam splitter in the interferometer. 3-5) At every step in the algorithm, we null a successive diagonal of the updated $U$ matrix, by alternating between $T_{m,n}$ and $T^{-1}_{m,n}$ matrices, which corresponds to adding diagonal lines of beam splitters to the interferometer. $T_{m,n}$ (resp. $T^{-1}_{m,n}$) matrices of a given color cause the rows (resp. columns) $m$ and $n$, which are shown in the same color, to mix, and null the corresponding element of that color in $U$. It is clear from this process that once a matrix element has been nulled, no subsequent operation can modify it. 6) After step 5, $U$ is a lower triangular matrix, which by virtue of its unitarity must be diagonal. As explained in the main text, we can then write $U$ in the way shown here, which by construction exactly corresponds to the desired interferometer.}
\label{fig2}
\end{figure*}

In this section, we present an analytical method of calculating the values of the beam splitter elements $T_{m,n}$ in our design. Beyond its practical utility in providing a recipe for programming such interferometers, the existence of this method serves to prove that our design is capable of implementing universal interferometric transformations. Our decomposition method relies on two important properties of the $T_{m,n}$ matrices. Firstly, for any given unitary matrix $U$, there are specific values of $\theta$ and $\phi$ that makes any target element in row $m$ or $n$ of matrix $T_{m,n}U$ zero, as per Reck \textit{et al} \cite{reck1994experimental}. We will refer to this process as nulling that element of $U$, and will still refer to the modified matrix after this operation as $U$. Secondly, we note that any target element in column $n$ or $m$ of $U$ can also be nulled by multiplying $U$ from the right by a $T^{-1}_{m,n}$ matrix.

We have constructed a simple algorithm, illustrated in figure \ref{fig2} for the $5 \times 5$ case, that enables us to synthesize an interferometer of arbitrary size consisting of concatenated $2 \times 2$ beam splitters of the kind given in equation \ref{generaldecompose}. We null elements of $U$ one by one in such a way that every $T_{m,n}$ and $T^{-1}_{m,n}$ matrix used in the process completely determines both the reflectivity and phase shift of one beam splitter and phase shifter. The protocol consists of nulling successive diagonals of $U$, in such a way that the sequence of $T_{m,n}$ and $T^{-1}_{m,n}$ matrices used both corresponds to the desired order of beam splitters in the interferometer, and guarantees that nulled elements of $U$ are not affected by subsequent operations. By construction, every nulled diagonal in the matrix corresponds to one diagonal line of beam splitters through the interferometer. By alternating between the left- and right-hand sides of the interferometer in our design, and thus between $T_{m,n}$ and $T^{-1}_{m,n}$ matrices, we fulfil the condition that no nulled element of $U$ is subsequently modified.

\begin{figure*}
\includegraphics[width=17cm]{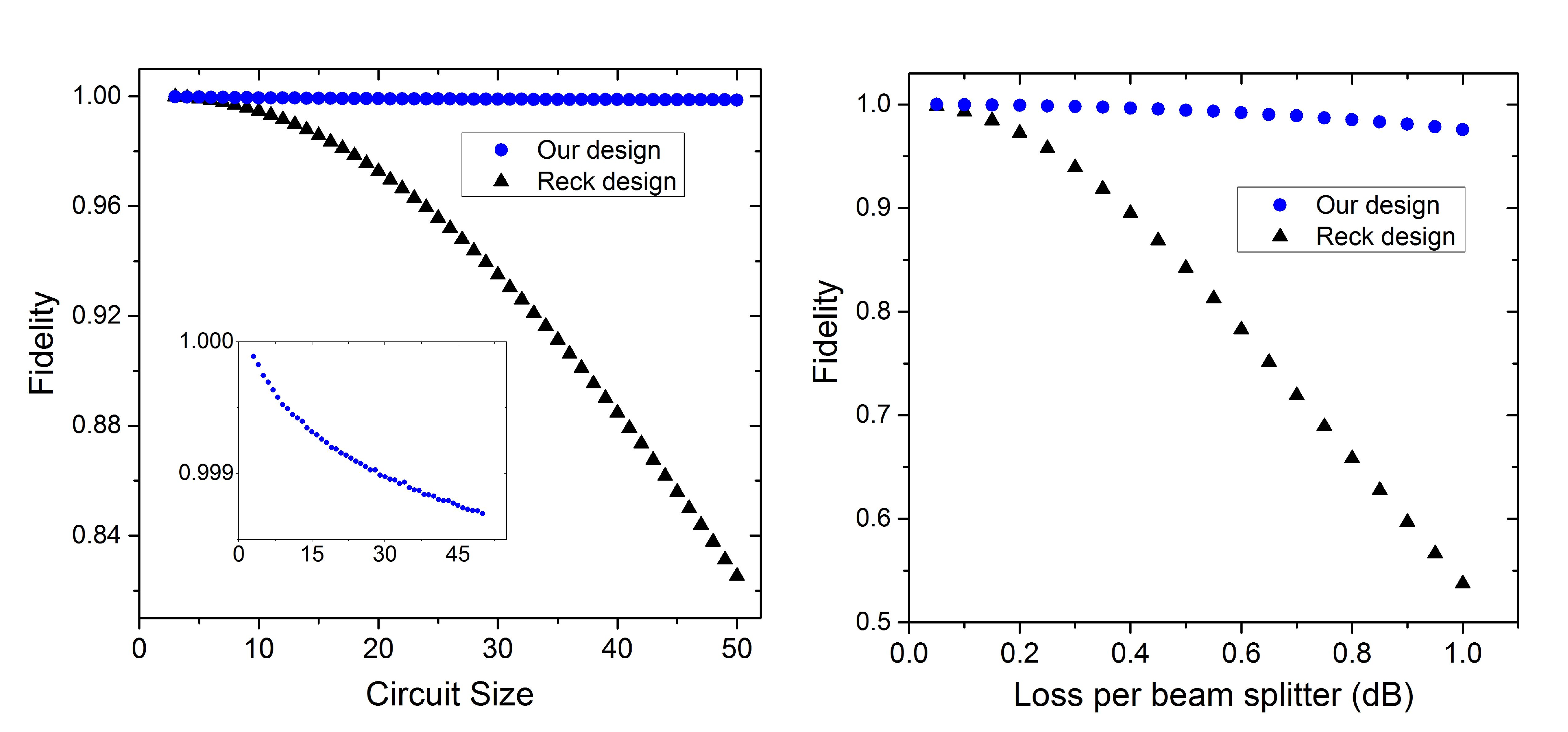}
\caption{Left: Average fidelity for an interferometer with a constant loss of 0.2~dB per beam splitter (as in the universal multiport interferometer in \cite{carolan2015universal}) for interferometers built according to the Reck design (black) and our design (blue), for different interferometer sizes. Inset: close-up of the fidelity in our design. Right: Fidelity as a function of loss, for interferometers implementing $20 \times 20$ transformations. We see from our results that our design is much more loss-tolerant than the Reck design, and maintains high fidelity with the target unitary even in the case of high loss. This is because mismatched path lengths in the Reck design causes loss to severely affect the resulting interference.}
\label{loss}
\end{figure*}

At the end of the decomposition process, we obtain the following expression for a $5 \times 5$ matrix:

\begin{align}
T_{4,5}T_{3,4}T_{2,3}T_{1,2} T_{4,5} T_{3,4} U T^{-1}_{1,2} T^{-1}_{3,4} T^{-1}_{2,3} T^{-1}_{1,2} = D
\end{align}

\noindent
where $D$ is a diagonal matrix as in equation \ref{generaldecompose}. This can be rewritten as:

\begin{align}
U = T^{-1}_{3,4} T^{-1}_{4,5} T^{-1}_{1,2}T^{-1}_{2,3} T^{-1}_{3,4} T^{-1}_{4,5} D T_{1,2} T_{2,3} T_{3,4} T_{1,2}
\end{align}

It is easy to demonstrate that, if $D$ consists of single-mode phase-shifts, then for any $T^{-1}_{m,n}$ matrix one can find a matrix $D'$ of single-mode phases and a matrix $T_{m,n}$ such that $T^{-1}_{m,n} D = D'T_{m,n}$. The previous equation can therefore be rewritten as:

\begin{align}
U  = D' T_{3,4} T_{4,5} T_{1,2} T_{2,3} T_{3,4} T_{4,5} T_{1,2} T_{2,3} T_{3,4} T_{1,2}
\label{finaleq}
\end{align}

\noindent
which, mirroring equation \ref{generaldecompose}, completes our decomposition. 

By construction, equation \ref{finaleq} physically corresponds to the multiport interferometer shown in figure \ref{fig2}, and the values of the $\theta$ and $\phi$ of the $T_{m,n}$ matrices in this equation determine the values of the beam splitters and phase shifts that must be programmed to implement $U$. This decomposition principle can be generalised to any $N$, and an explicit general algorithm is given in the supplementary information. We also note that this algorithm can be used to inform the design of fixed interferometric circuits, such as those demonstrated in \cite{bentivegna2015experimental,spagnolo2014experimental}, in which the same arrangement of beam splitters was used to provide specific instances of random interference.
 
\subsection{Loss tolerance}

Optical loss is unavoidable in realistic interferometers, and finding methods to mitigate its effects is an integral part of any photonic scheme. In the following, we study the tolerance of multiport interferometers built according to our decomposition to loss, and compare their performance to interferometers built and programmed according to the Reck design. 

We first distinguish between two types of loss. Balanced loss in a multiport interferometer, in which every path through the interferometer experiences the same loss, preserves the target interference to within an overall scaling factor. This is generally acceptable for applications in the classical domain, such as optical switching or microwave photonics.  In the quantum domain, although loss severely affects scalability of quantum experiments, post-selection can in some situations be used to recover the desired interference pattern. We note that propagation loss in an interferometer is expected to contribute to balanced loss, since every physical path length in an interferometer must be matched to within the coherence length of the input light to maintain high-fidelity interference. However, propagation loss must therefore be proportional to the longest path through the interferometer (i.e. the optical depth), so interferometers built according to our design will suffer from only about half the propagation loss of an interferometer built according to the Reck design.

Unbalanced loss, where different paths through the interferometer experience different loss, can be difficult to characterize and, critically, can result in a poor fidelity to the intended operation \cite{metcalf2013multiphoton,metcalf2014quantum,bonneau2012quantum,peruzzo2011multimode}. Unequal losses between paths in the interferometer are typically caused by beam splitters, which are unavoidably lossy due to additional bending losses and scattering. To compare the tolerance of multiport interferometers to unbalanced loss caused by beam splitters, we adopt the following procedure. For a given $N$, we generate 500 random unitary matrices \cite{note}, implement our decomposition, add loss to both outputs of all the resulting beam splitters, and compare the fidelities in the overall transformations. We use a simple loss model that assumes equal insertion loss for every beam splitter, and we quantify the fidelity of the transformation implemented by a lossy $N \times N$ experimental interferometer, $U_{exp}$, to the intended transformation $U$ using the following metric:

\begin{align}
F(U_{exp},U) = \left| \frac{\text{tr}(U^\dagger U_{exp})}{\sqrt{N \text{tr}(U_{exp}^\dagger U_{exp})}} \right|^2
\end{align}

\noindent
which corresponds to a standard fidelity measure, normalized so that we do not distinguish between matrices that differ by only a constant multiplicative factor. This allows us to focus on unbalanced loss instead of balanced loss in our simulations.

Figure \ref{loss} shows our simulation results, for both a fixed loss and varying interferometer sizes, and for a fixed interferometer size and varying loss. We conclude that interferometers that implement our design are significantly more tolerant to unbalanced loss than those implementing the Reck design. This is because, in the Reck design, different paths through the interferometer go through different numbers of beam splitters, so they all experience different loss and the resulting interference is degraded. In our design, the path lengths are better matched, so equally distributed loss within the interferometer does not strongly affect the resulting interference. We note that whereas unbalanced loss can be compensated for in the Reck design by adding loss to shorter paths, for example by adding beam splitters to the shorter paths in the interferometer as proposed by Miller \cite{miller2015perfect}, this is inefficient and it is better to start with a fundamentally loss-resistant interferometer.

\subsection{Conclusion}

In conclusion, we have demonstrated a design for universal multiport interferometers that outperforms the design proposed by Reck \textit{et al} in several respects. Our design is programmed using a new method for decomposing unitary matrices into a sequence of beam splitters, is almost twice as compact and, significantly, not only suffers less propagation loss but is more loss-tolerant than the previous design.

We expect that our compact and loss tolerant design for fully programmable universal mulitport interferometers will play an important role in the development of optical processors for both classical and quantum applications. Furthermore, we anticipate that our matrix decomposition method will be of use in its own right for other systems which use mathematical structures analogous to beam splitters and phase shifters, such as ion traps \cite{shen2014scalable} and some architectures for superconducting circuits \cite{peropadre2015spin,peropadre2015microwave}. 

\section*{Acknowledgements}
This work was supported by the European Research Council, the UK Engineering and Physical Sciences Research Council (project EP/K034480/1 and the Networked 
Quantum Information Technology Hub), and by the European Commission (H2020-FETPROACT-2014 grant QUCHIP).

\section*{Author Contributions}
\footnotesize{W.R.C. and P.C.H. demonstrated the matrix decomposition method presented here. W.R.C did the numerical simulations and wrote the manuscript with contributions from all authors. B.J.M. conceived the project, and W.S.K, and I.A.W. supervised it.}

\section*{Additional Information}
\subsection*{Competing financial interests}
The authors have jointly applied for a patent for the work presented in this paper.
\subsection*{Supplementary Information}
Accompanies this paper

\bibliographystyle{ieeetr}
\bibliography{ArXivVersion}

\begin{thebibliography}{10}

\bibitem{ref1}
``Birth of the programmable optical chip,'' {\em Nat Photon}, vol.~10,
  pp.~1--1, Jan 2016.
\newblock Editorial.

\bibitem{capmany2016programmable}
J.~Capmany, I.~Gasulla, and D.~Perez, ``Microwave photonics: The programmable
  processor,'' {\em Nat Photon}, vol.~10, pp.~6--8, Jan 2016.
\newblock News and Views.

\bibitem{chen2011photonic}
L.-N. Chen, E.~Hall, L.~Theogarajan, and J.~Bowers, ``Photonic switching for
  data center applications,'' {\em IEEE Photonics journal}, vol.~3, no.~5,
  pp.~834--844, 2011.

\bibitem{stabile2016integrated}
R.~Stabile, A.~Albores-Mejia, A.~Rohit, and K.~A. Williams, ``Integrated
  optical switch matrices for packet data networks,'' {\em Microsystems \&
  Nanoengineering}, vol.~2, 2016.

\bibitem{reck1994experimental}
M.~Reck, A.~Zeilinger, H.~J. Bernstein, and P.~Bertani, ``Experimental
  realization of any discrete unitary operator,'' {\em Physical Review
  Letters}, vol.~73, no.~1, p.~58, 1994.

\bibitem{carolan2015universal}
J.~Carolan, C.~Harrold, C.~Sparrow, E.~Martin-L\'{o}pez, N.~J. Russell, J.~W.
  Silverstone, P.~J. Shadbolt, N.~Matsuda, M.~Oguma, M.~Itoh, G.~D. Marshall,
  M.~G. Thompson, J.~C.~F. Matthews, T.~Hashimoto, J.~L. O'Brien, and A.~Laing,
  ``Universal linear optics,'' {\em Science}, 2015.

\bibitem{tanizawa2015ultra}
K.~Tanizawa, K.~Suzuki, M.~Toyama, M.~Ohtsuka, N.~Yokoyama, K.~Matsumaro,
  M.~Seki, K.~Koshino, T.~Sugaya, S.~Suda, {\em et~al.}, ``Ultra-compact
  32$\times$ 32 strictly-non-blocking \uppercase{S}i-wire optical switch with
  fan-out \uppercase{LGA} interposer,'' {\em Optics express}, vol.~23, no.~13,
  pp.~17599--17606, 2015.

\bibitem{harris2015bosonic}
N.~C. Harris, G.~R. Steinbrecher, J.~Mower, Y.~Lahini, M.~Prabhu,
  T.~Baehr-Jones, M.~Hochberg, S.~Lloyd, and D.~Englund, ``Bosonic transport
  simulations in a large-scale programmable nanophotonic processor,'' {\em
  arXiv preprint arXiv:1507.03406}, 2015.

\bibitem{gaylor1990integrated}
T.~K. Gaylor, E.~I. Verriest, and M.~M. Mirsalehi, ``Integrated optical givens
  rotation device,'' Aug.~21 1990.
\newblock US Patent 4,950,042.

\bibitem{miller2013self}
D.~A. Miller, ``Self-configuring universal linear optical component,'' {\em
  Photonics Research}, vol.~1, no.~1, pp.~1--15, 2013.

\bibitem{miller2015perfect}
D.~A. Miller, ``Perfect optics with imperfect components,'' {\em Optica},
  vol.~2, no.~8, pp.~747--750, 2015.

\bibitem{alber2003quantum}
G.~Alber, T.~Beth, M.~Horodecki, P.~Horodecki, R.~Horodecki, M.~R{\"o}tteler,
  H.~Weinfurter, R.~Werner, and A.~Zeilinger, {\em Quantum Information: An
  introduction to basic theoretical concepts and experiments}, vol.~173.
\newblock Springer, 2003.

\bibitem{bentivegna2015experimental}
M.~Bentivegna, N.~Spagnolo, C.~Vitelli, F.~Flamini, N.~Viggianiello,
  L.~Latmiral, P.~Mataloni, D.~J. Brod, E.~F. Galv{\~a}o, A.~Crespi, {\em
  et~al.}, ``Experimental scattershot boson sampling,'' {\em Science Advances},
  vol.~1, no.~3, p.~e1400255, 2015.

\bibitem{spagnolo2014experimental}
N.~Spagnolo, C.~Vitelli, M.~Bentivegna, D.~J. Brod, A.~Crespi, F.~Flamini,
  S.~Giacomini, G.~Milani, R.~Ramponi, P.~Mataloni, {\em et~al.},
  ``Experimental validation of photonic boson sampling,'' {\em Nature
  Photonics}, vol.~8, no.~8, pp.~615--620, 2014.

\bibitem{metcalf2013multiphoton}
B.~J. Metcalf, N.~Thomas-Peter, J.~B. Spring, D.~Kundys, M.~A. Broome, P.~C.
  Humphreys, X.-M. Jin, M.~Barbieri, W.~S. Kolthammer, J.~C. Gates, {\em
  et~al.}, ``Multiphoton quantum interference in a multiport integrated
  photonic device,'' {\em Nature communications}, vol.~4, p.~1356, 2013.

\bibitem{metcalf2014quantum}
B.~J. Metcalf, J.~B. Spring, P.~C. Humphreys, N.~Thomas-Peter, M.~Barbieri,
  W.~S. Kolthammer, X.-M. Jin, N.~K. Langford, D.~Kundys, J.~C. Gates, B.~J.
  Smith, P.~G.~R. Smith, and I.~A. Walmsley, ``Quantum teleportation on a
  photonic chip,'' {\em Nature Photonics}, vol.~8, no.~10, pp.~770--774, 2014.

\bibitem{bonneau2012quantum}
D.~Bonneau, E.~Engin, K.~Ohira, N.~Suzuki, H.~Yoshida, N.~Iizuka, M.~Ezaki,
  C.~M. Natarajan, M.~G. Tanner, R.~H. Hadfield, {\em et~al.}, ``Quantum
  interference and manipulation of entanglement in silicon wire waveguide
  quantum circuits,'' {\em New Journal of Physics}, vol.~14, no.~4, p.~045003,
  2012.

\bibitem{peruzzo2011multimode}
A.~Peruzzo, A.~Laing, A.~Politi, T.~Rudolph, and J.~L. O'Brien, ``Multimode
  quantum interference of photons in multiport integrated devices,'' {\em
  Nature communications}, vol.~2, p.~224, 2011.

\bibitem{note}
These random unitary matrices were generated using the QR decomposition of
  random matrices, following the code developed by Toby Cubitt, available on
  \url{www.dr-qubit.org/matlab.php}.

\bibitem{shen2014scalable}
C.~Shen, Z.~Zhang, and L.-M. Duan, ``Scalable implementation of boson sampling
  with trapped ions,'' {\em Physical review letters}, vol.~112, no.~5,
  p.~050504, 2014.

\bibitem{peropadre2015spin}
B.~Peropadre, A.~Aspuru-Guzik, and J.~J. Garcia-Ripoll, ``Spin models and boson
  sampling,'' {\em arXiv preprint arXiv:1509.02703}, 2015.

\bibitem{peropadre2015microwave}
B.~Peropadre, G.~G. Guerreschi, J.~Huh, and A.~Aspuru-Guzik, ``Microwave boson
  sampling,'' {\em arXiv preprint arXiv:1510.08064}, 2015.

\bibitem{mower2015high}
J.~Mower, N.~C. Harris, G.~R. Steinbrecher, Y.~Lahini, and D.~Englund,
  ``High-fidelity quantum state evolution in imperfect photonic integrated
  circuits,'' {\em Physical Review A}, vol.~92, no.~3, p.~032322, 2015.

\bibitem{thomas2011integrated}
N.~Thomas-Peter, N.~K. Langford, A.~Datta, L.~Zhang, B.~J. Smith, J.~B. Spring,
  B.~J. Metcalf, H.~B. Coldenstrodt-Ronge, M.~Hu, J.~Nunn, {\em et~al.},
  ``Integrated photonic sensing,'' {\em New Journal of Physics}, vol.~13,
  no.~5, p.~055024, 2011.

\bibitem{thomas2012quantum}
N.~Thomas-Peter, {\em Quantum enhanced precision measurement and information
  processing with integrated photonics}.
\newblock PhD thesis, Balliol College, University of Oxford, 2012.

\end{thebibliography}

\pagebreak
\onecolumngrid

\section{Supplementary Information}

\subsection{Characterizing a realistic universal multiport interferometer}

Programming a universal multiport interferometer using our procedure requires a preliminary full characterization of its beam splitters and phase shifters. This is a simple procedure, similar in spirit to that proposed by Mower \textit{et al} \cite{mower2015high}, and only has to be done once, provided that there is no long-term drift of the optical properties of the interferometer. 

At every step in the process, we choose a path through the interferometer which can be broken by setting a single beam splitter in the path to full transmission. We then input light into that path, and scan through the reflectivity of that beam splitter while monitoring the output. This allows us to characterise that beam splitter. We then set it to be fully transmissive, and move on to a different path until every beam splitter has been characterised and the interferometer implements the identity to within single-mode phase shifts.

Individual phase shifters can then be characterized by creating simple interfering paths through the interferometer, and modulating the phase shifters in those paths. Every interfering path consists of several phase shifters, but since there are many more possible interfering paths than phase shifters, the phase shifters can still be individually characterized. We note that the phase shifters at the input of the interferometer cannot be individually characterised in this way, but these are typically not relevant for most applications.

The preceding protocol assumes that the beam splitters can perfectly implement the identity. This is typically not the case for real interferometers, where small amounts of light will leak through. However, the approach proposed by Mower \textit{et al} to overcome this problem also works for our design. This light can be isolated and removed from the characterisation process by varying the reflectivities of the beam splitters not along the path being broken, in such a way that the spurious light can be identified in the Fourier transform of the output signal.

\subsection{General decomposition procedure}

\begin{figure}
\includegraphics[scale=0.3]{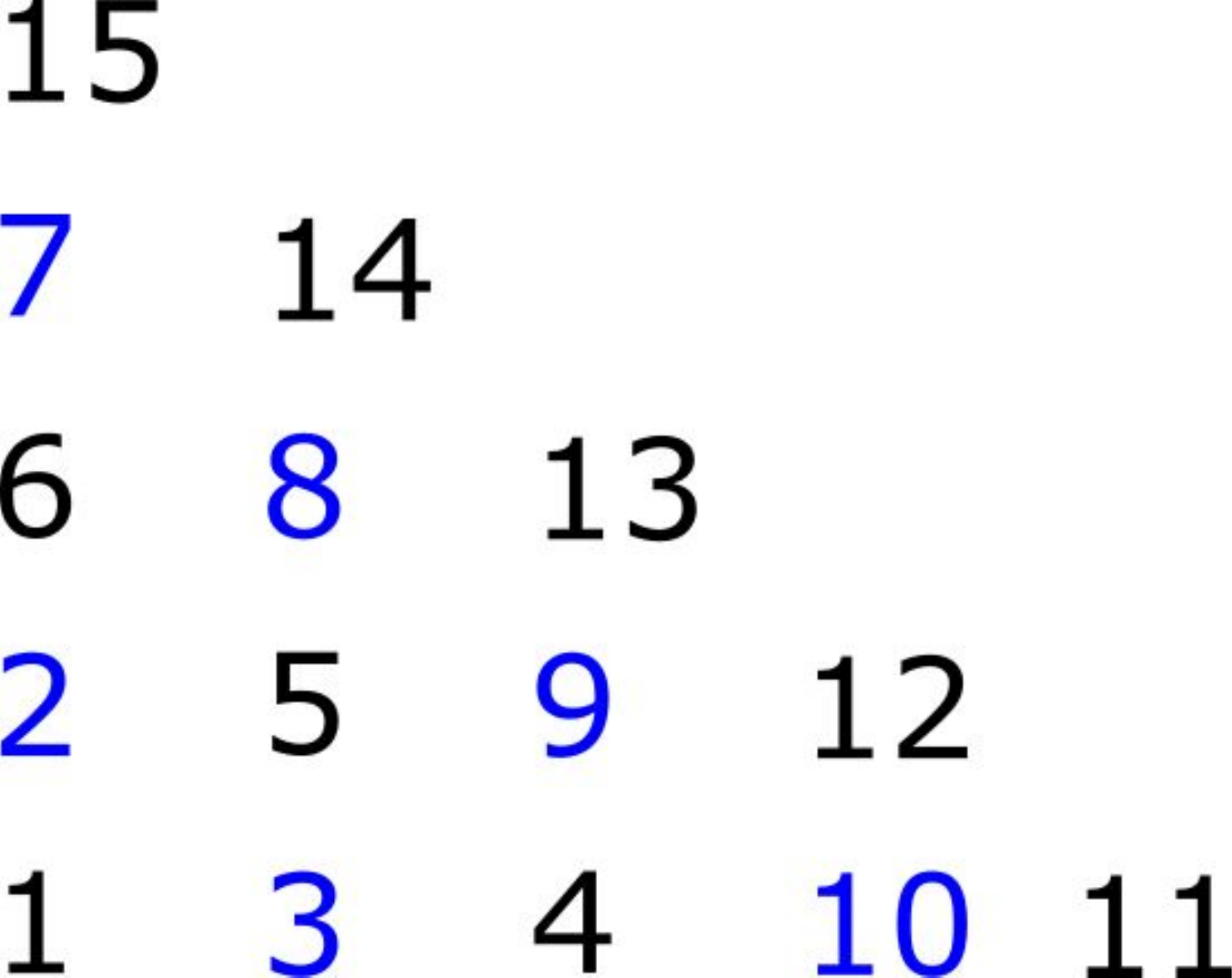}
\caption{Illustration of the order in which matrix elements of a unitary matrix $\hat{U}$ are nulled. The first element to be nulled is at the bottom left of the matrix. The following elements are then nulled in consecutive diagonals. A black element located in column $i$ is nulled with a $T^{-1}_{i,i+1}$ matrix, and a blue element located in row $i$ is nulled with a $T_{i-1,i}$ matrix.}
\label{fig1_SI}
\end{figure}

The unitary matrix decomposition procedure presented in the main text can easily be generalised to any $N \times N$ unitary matrix. Elements of $\hat{U}$ are consecutively nulled using $T_{m,n}$ or $T^{-1}_{m,n}$ matrices, which physically correspond to beam splitters in the final interferometer, in the pattern shown in figure \ref{fig1_SI}.

The algorithm that implements the decomposition is the following:\\ \\
for $i$ from 1 to $N-1$ \\ \\
\hspace*{1cm} if $i$ is odd  \\
\hspace*{2cm} for $j = 0$ to $i-1$\\
\hspace*{3cm} find a $T^{-1}_{i-j,i-j+1}$ matrix that nulls element $(N-j,i-j)$ of $\hat{U}$\\
\hspace*{3cm} update $\hat{U} = \hat{U}T^{-1}_{i-j,i-j+1}$.\\
\hspace*{2cm} end for\\ \\
\hspace*{1cm} else if $i$ is even  \\
\hspace*{2cm} for $j = 1$ to $i$\\
\hspace*{3cm} find a $T_{N+j-i-1,N+j-i}$ matrix that nulls element $(N+j-i,j)$ of $\hat{U}$\\
\hspace*{3cm} update $\hat{U} = T_{N+j-i-1,N+j-i}\hat{U}$\\
\hspace*{2cm} end for\\
\hspace*{1cm} end if \\ \\
end for \\ \\

After this decomposition procedure, we obtain the following expression:

\begin{align*}
\left(\prod_{(m,n) \in S_L} T_{m,n}\right) \hat{U} \left(\prod_{(m,n) \in S_R} T^{-1}_{m,n}\right) = D
\end{align*}

\noindent
where $D$ is a diagonal matrix corresponding to single-mode phases, and $S_L$ and $S_R$ are the respective orderings of the $(m,n)$ indices for the $T_{m,n}$ or $T^{-1}_{m,n}$ matrices yielded by our decomposition. This can be rewritten as:

\begin{align*}
 \hat{U}  = \left(\prod_{(m,n) \in S^T_L} T^{-1}_{m,n}\right) D \left(\prod_{(m,n) \in S^T_R} T_{m,n}\right)
\end{align*}

We can then find a matrix $D'$ and $T_{m,n}$ matrices such that the previous equation can be re-written as:

\begin{align*}
\hat{U} = D' \left(\prod_{(m,n) \in S} T_{m,n}\right)
\end{align*}

\noindent
where $S$ is, by construction, the order of beam splitters corresponding to the desired circuit. This completes our decomposition.

\subsection{Error}

In a realistic interferometer, there will always be some finite error when setting the values for the phases, even after the characterisation procedure described above. Furthermore, imperfections in the beam splitters will always make it difficult to reach perfect transmission or reflection, although we do note that concatenating imperfect beam splitters to create one perfect beam splitter \cite{thomas2011integrated,thomas2012quantum, miller2015perfect} overcomes this problem, at the cost of adding beam splitters. These errors will affect the circuit fidelity in both our design and in the Reck design.

However, the overall error in the interferometer caused by these individual errors depends on the total number of beam splitters, and the layout of the interferometer only affects how that error is distributed among the output ports. Therefore, the average error is roughly equal in the Reck design and in our design, although it is more evenly distributed along the output modes in our design.

\end{document}